# Stopping power of a heterogeneous Warm Dense Matter

**David Casas[1,2] , Alexander A. Andreev[2], Matthias Schnürer[2], Manuel D. Barriga-Carrasco[1], Roberto Morales[1] and Luis González-Gallego[1]**

*[1]E.T.S.I. Industriales, Universidad de Castilla-La Mancha, E-13071 Ciudad Real, Spain*

*[2]Max Born Institute, Max Born Str. 2a D-12489, Berlin, Germany*

Corresponding author:

David Casas

E.T.S.I. Industriales, Universidad de Castilla-La Mancha, 13071, Ciudad Real, Spain.

E-mail: david.casas@uclm.es; tel. (34) 926 295 300 - Ext. 96688

Number of manuscript pages = 16

Number of figures = 9

# Stopping power of a heterogeneous Warm Dense Matter

## Abstract

The stopping power of Warm Dense Matter (WDM) is estimated by means of the individual contributions of free electrons and bound electrons existing in this special kind of matter, located between classical and degenerate plasmas. For free electrons, the dielectric formalism, well described in our studies, is used to estimate the free electron stopping power. For bound electrons, the mean excitation energy of ions is used. Excitation energies are obtained through atomic calculations of the whole atom or, shell by shell in order to estimate their stopping power. Influence of temperature and density is analyzed in case of an impinging projectile. This influence becomes important for low projectile velocities and is negligible for high ones. Using free and bound electron analysis, the stopping power of an extended WDM is inferred from a dynamical calculation of energy transferred from the projectile to the plasma, where the stopping range is calculated. Finally, this theoretical framework is used to study a typical plasma density profile of a WDM heated by lasers.

**Introduction**

In the last decades, the interaction of ion beams with plasmas has been widely studied for light and heavy ions (Ziegler, 1999; Zwicknagel et al., 1999). For the last ones, the charge state of the projectile has to be taken into account, and several experiments have been performed in order to analyze the evolution of its charge inside the plasma and the energy loss of ions (Gardes et al., 1992; Couillaud et al., 1994; Hoffmann et al., 1990; Chabot et al., 1995; Jacoby et al., 1995). In the case of protons, it is not necessary to study the charge state, the stopping power depends only on medium characteristics. Energy loss of proton beams in plasmas has been measured during the last years (Golubev et al., 1998; Mintsev et al., 1999; Golubev et al., 2001; Shibata et al., 2001) covering proton kinetic energies from 350 keV to 6 MeV and electron densities from $10^{17}$ cm$^{-3}$ to $10^{19}$ cm$^{-3}$.

During last years, laser-accelerated protons has been established as a new technique that allows to obtain proton beams with energies from several to tens of MeV. Compared to classical electromagnetic accelerators method, the laser-driven method has specific advantages which are related to the acceleration process at ultrafast timescale and at relativistic intensity. These proton and ion beams have an ultralow longitudinal and transversal emittance, are emitted from a small source at ultra-short temporal duration and have an energy distribution from several to tens of MeV (Daido et al., 2012). Target Normal Sheath Acceleration (TNSA) is one commonly used method to accelerated protons and ions using high power laser pulses. Here the plasma, created from a thin target using a laser pulse, has a non-neutral area, near the expansion front, called plasma sheath. In this region the gradient of the electrostatic field causes the acceleration of electrons, protons and ions, well described in this model (Macchi et al., 2013).

Laser accelerated protons can be utilized as probes in fast dynamical phenomena investigating plasma density and electromagnetic fields space- and time- resolved. The proton beam is deflected by electric and magnetic fields, while stopping and scattering effects are due to plasma density (Mackinnon et al., 2004; Borghesi et al., 2002). Using the "proton streak deflectometry" technique, time resolution and spatial geometry in one dimension is resolved together in order to show the ultrafast and transient field front associated with the TNSA starting regime where electromagnetic effects come into play (Sokollik et al., 2008; Abicht et al., 2014).

Laser generated proton beams are applied to heat up solid density matter via isochoric heating and to create Warm Dense Matter (WDM) states, i.e. matter at 1-10 times solid density and temperatures up to 100 eV. Using a close distance to the protons source and the proton pulse duration in the range of picoseconds causes an isochoric target heating in timescale shorter than the hydrodynamic plasma expansion timescale (Patel et al., 2003; Dyer et al., 2008). Thus the Equation of State (EOS) of WDM can be studied in this temporal window (Mancic et al., 2010; Pelka et al., 2010).

The stopping power of a plasma can be studied in two cases. The first one used a fully ionized plasma, when the ions have been stripped off all their electrons moving freely in the plasma. In this case, many dielectric functions and methods have been developed (Lindhard, 1954; Peter and Meyer-ter-Vehn, 1991; Mermin, 1970; Barriga-Carrasco, 2010). The second one deals with a partially ionized plasma, where the ions remain some of their electrons. In such a plasma the electron stopping consists of two parts: one is due to free electrons being estimated with the same methods as the fully ionized plasma and the second one is due to bound electrons that contribute to the plasma stopping power with excitation and/or ionization electron processes (Garbet et al., 1987).

For the first case, the Random Phase Approximation (RPA) dielectric function is used. This way the effect of the incident charged particle as a perturbation is considered and energy transfer to target is proportional to the square of its charge. The linear response theory is usually applicable for high-velocity projectiles and in the weak coupling of an electron gas. Then the slowing down of impinging ions is simplified to a treatment of the properties of the medium only, and a linear description of these properties may be then applied (Barriga-Carrasco, 2010; Barriga-Carrasco and Casas, 2013). Using the dielectric formalism, analytical formulas with few parameters for classical plasmas have been proposed (Peter and Meyer-ter-Vehn, 1991). These expressions work well for classical plasmas, i.e. high temperatures and low densities but give inaccurate results for degenerate plasmas (low temperatures and high densities). Other interpolation formulas have been proposed for plasmas of any degeneracy by means of coefficients (Maynard and Deutsch, 1985). The use of many mathematical integrals in the calculation results in a complex calculation of these dielectric functions, as shown in the next section. To join a fast calculation of analytical formulas with the accuracy of a full dielectric function that works with plasmas of all degeneracies, e.g. Random Phase Approximation (RPA), an alternative method uses interpolation from a complete database of stopping power of a wide range of plasma parameters (Barriga-Carrasco, 2013).

In the case of bound electron stopping power, mean excitation energies of atoms or ions, $I$, can be obtained with several methods like Hartree-Fock (HF), Oscillator Strength (OS), Local Plasma Approximation (LPA), or atomic potentials in an independent particle model (Casas et al., 2013; Lindhard and Scharff, 1953; Garbet et al., 1987). In the first two methods, $I$ is obtained for every atomic shell, where for the two last methods, $I$ is a global quantity for the whole atom or ion.

There is a great interest to study the influence of stopping power of a plasma density profile in a real case when the plasma is created from interaction of a laser pulse with a thin target and a plasma gradient is modified by the laser prepulse. Instead of a constant plasma density distribution, there will be an extended distribution with plasma density gradients. In this case an analytical formula that represents this density profile is used in order to study the interaction of the main pulse with the plasma and the energy loss of proton beam (Andreev et al., 2009).

In the following section "Theoretical methods", stopping power expressions and formulas for free and bound electrons will be given. Afterwards, in section "Calculations" it will be explained how the energy loss and the stopping range of the passing ion beam through the plasma is obtained. Then this method is used to study the influence of the plasma density profile concerning energy loss, and temperature distribution in WDM. Finally, these results will be discussed in the section "Conclusions".

## Theoretical methods

### Free electron stopping

In the classical work of Peter and Meyer-ter-Vehn (1991), an analytic approximation of stopping power of free electrons for arbitrary projectile velocities is deduced inside the theoretical framework of Vlasov linear theory of stopping power. There, the plasma parameter

$$Z = \frac{Zeff}{n_e \lambda_D^3},$$ (1)

where $Z_{eff}$ is the projectile effective charge, $n_e$ is the electron density and $\lambda_D$ is the Debye length. This parameter separates in two regions the plasma by means of their degeneracy, for classical plasmas is $Z < 1$, while for degenerate plasmas is $Z \geq 1$. In the study of quantum plasmas, other degeneracy parameter (Arista and Brandt, 1984) is used:

$$D = \frac{E_F}{k_B T},$$ (2)

where $E_F$ is the Fermi energy. Plasma is degenerate for $D \gg 1$ and non-degenerate for $D \ll 1$. The two regions for both parameters are represented in a density-temperature map showed in Figure 1.

-----------------------------Figure 1--------------------------------

For $Z < 1$, an analytic formula was proposed by Peter and Meyer-ter-Vehn (1991):

$$Sp = \left[ \frac{Z_{eff} e \omega_p}{v_p} \right]^2 \left[ G\left( \frac{v_p}{\sqrt{k_B T/m}} \right) \ln\left( k_{max} \lambda_D \right) + H\left( \frac{v_p}{\sqrt{k_B T/m}} \right) \ln\left( \frac{v_p}{\sqrt{k_B T/m}} \right) \right],$$ (3)

where

$$G\left( v_p \right) = \text{erf} \left[ \frac{v_p}{\sqrt{2}} \right] - \left[ \frac{2}{\pi} \right]^{1/2} v_p \exp\left( -v_p^2/2 \right),$$ (4)

$$H\left(v_p\right) = -\frac{v_p^3}{3\sqrt{2\pi}\ln v_p}\exp\left(-v_p^2/2\right) + \frac{v_p^4}{v_p^4 + 12}, \tag{5}$$

Here, $v_p$ is projectile velocity, $e$ is electron charge, $\omega_p$ is plasma frequency, $T$ is temperature, $k_B$ is Boltzmann constant, and $k_{max}$ a cut-off parameter that is chosen between classical and quantum limits (Arista and Brandt, 1984).

$$\begin{cases} k_{max\,Cl} = m\left(v_p^2 + 2k_B T/m\right)\big/\big|Z_{eff}\big|e^2 \\ k_{max\,Qu} = 2mv_p/\hbar \end{cases} \tag{6}$$

The equations from (3) to (6) have been simplified by Volpe *et al.* (2011) for the case of protons ($Z_{eff} = 1$), where is given the stopping power as a function of proton energy instead of projectile velocity. Eq. (3) is used for pure classical plasmas. However, for intermediate degeneration or quantum plasmas, it is necessary to perform calculations using the integration of the dielectric function given in this work.

The study of stopping power for plasmas of all degeneracy degrees can be performed using the RPA dielectric function (DF), which consists of considering the effect of the incident particle as a perturbation, so that the energy loss was proportional to the square of its charge. Then slowing-down was simplified to a treatment of the properties of the medium only, and a linear description of these properties may then be applied.

This DF is developed in terms of the wave number $k$ and of the frequency $\omega$ provided by a consistent quantum mechanical analysis. The RPA analysis yields to the expression (Lindhard, 1954):

$$\varepsilon_{\mathrm{RPA}}(k,\omega) = 1 + \frac{1}{\pi^2 k^2}\int d^3 k' \frac{f(\vec{k}+\vec{k}') - f(\vec{k}')}{\omega + i\upsilon - (E_{\vec{k}+\vec{k}'} - E_{\vec{k}'})} \ , \tag{7}$$

where $E_{\vec{k}} = k^2/2$. The temperature dependence is included through the Fermi-Dirac function

$$f(\vec{k}) = \frac{1}{1 + \exp[\beta(E_k - \mu)]} \tag{8}$$

being $\beta = 1/k_B T$ and $\mu$ the chemical potential of the plasma. In this part of the analysis we assume the absence of collisions so that the collision frequency tends to zero, $\upsilon \to 0$ .

Analytic RPA DF for plasmas at any degeneracy can be obtained directly from Eq. (7) (Gouedard and Deutsch, 1978; Arista and Brandt, 1984)

$$\varepsilon_{\mathrm{RPA}}(k,\omega) = 1 + \frac{1}{4z^3\pi k_F}[g(u+z) - g(u-z)], \tag{9}$$

where g(x) corresponds to

$$g(x) = \int_0^\infty \frac{ydy}{\exp(Dy^2 - \beta\mu) + 1} \ln\left(\frac{x+y}{x-y}\right) \qquad (10)$$

$u = \omega/kv_F$ and $z = k/2k_F$ are the common dimensionless variables (Lindhard, 1954). $D = E_F\beta$ is the degeneracy parameter and $v_F = k_F = \sqrt{2E_F}$ is Fermi velocity in a.u.

Finally, electronic stopping of free plasma electrons will be calculated in the dielectric formalism as

$$Sp(v_p) = \frac{2Z_{eff}^2}{\pi v_p^2} \int_0^\infty \frac{dk}{k} \int_0^{kv} d\omega\, \omega \operatorname{Im}\left[\frac{-1}{\varepsilon_{RPA}(k,\omega)}\right] \quad \text{(a.u.)} \qquad (11)$$

The calculation of the integrals in Eq. (11) could be difficult and computationally slow in some cases. A fast accurate method is to make an interpolation from a database (Barriga-Carrasco, 2013). In Figure 2, we can see the stopping power calculated from Eq. (3) and Eq. (11) for the same plasma conditions. Both stoppings are similar each to other.

----------------------------Figure 2--------------------------------

Stopping power dependence on temperature and electron density as function of projectile velocity is observed for low-medium velocities and vanishes for high velocities. In this last case, stopping power is a function approximately linear of electron density.

In order to illustrate these effects, we choose a region of classical plasmas to compare equations (3) and (11), because a complete study for all degeneracy is off the limits of this work.

The stopping power rises when the electron density grows, but not linearly. There are two dependencies in Eq. (3). One is linear, because $\omega_p = \sqrt{4\pi n_e}$, and the other one is logarithmic due to $\ln(k_{max}\lambda_D)$. When all the parameters remain constant except density, the Eq. (3) can be expressed as a density function:

$$Sp(n_e) = n_e\left[K_\alpha \ln\left(1/\sqrt{n_e}\right) + K_\beta\right] \qquad (12)$$

The consequence of Eq. (12) is that stopping power increases less than the growth of density, how we can see in Figure 3, where it is showed the stopping power ratios for different densities from $1 \cdot 10^{19}$ e⁻/cm³ to $1 \cdot 10^{22}$ e⁻/cm³. All of them are divided for the stopping corresponding to $1 \cdot 10^{19}$ e⁻/cm³ ($Sp_0$) and plotted as a function of proton energy in MeV. For high velocities and small densities variations inside the plasma, we can considerer stopping power as quasilinear function of density.

----------------------------Figure 3--------------------------------

In the case of temperature, if we calculate the stopping ratio for many temperatures with constant density of $1 \cdot 10^{19}$ e$^-$/cm$^3$ by means of the stopping for a temperature of 50 eV ($Sp_0$), the exponentials in equations (3), (4), and (5) show a decay exponential behavior with an asymptotic limit equal to 1, as it is shown in Figure 4. We can approximate this behavior with the exponential expression, where $R_0$ is equal to unity:

$$R_{Sp} = R_0 + ae^{-v_p/b} \qquad (13)$$

----------------------------Figure 4--------------------------------

Due to these results the stopping power calculations for slow projectiles in heterogeneous plasmas with high gradients in temperature and density will be very sensible to the theoretical methods used to estimate it (Gericke and Schlanges, 2003).

In the other hand for high velocities projectiles, we can approximate the stopping power as linear function of density.

Bound electron stopping

The stopping power of a cold gas or a plasma for an ion has been calculated many times by the well-known Bethe formula (Bethe, 1930):

$$Sp = \left[\frac{Z_{eff}e\omega_p}{v_p}\right]^2 \ln\left[\frac{2mv_p}{I}\right], \qquad (14)$$

where $I$ is the mean excitation energy, which averages all the exchanged energy in excitation and/or ionization processes between a swift charged particle and target bound electrons. $I$ can be easily calculated for every neutral atom as a function of atomic number $Z$, $I = 10.3Z$ (eV) , (Garbet et al., 1987). For ions, $I$ can be obtained solving Poisson´s equation of atomic electron density (Garbet, Deutsch et al. 1987) using analytic atomic potentials with the independent particle model (Green, Sellin et al. 1969). $I$, $K$ and $r^2$ are estimated using this method as:

$$I^2 = \frac{Z^4}{N^3}\frac{H^2}{6d^3}\frac{\left[1-\left(\frac{N}{6Z}\right)\left(1+\frac{1}{2H}\right)\right]}{F(\alpha)}, \qquad (15)$$

$$K = \frac{Z^3}{N^3}\frac{H}{2d}\left[Z-N\left(\frac{1}{6}+\frac{1}{12H}\right)\right], \qquad (16)$$

$$r^2 = \frac{6d^2}{H} F(\alpha), \tag{17}$$

where $Z$ is the atomic number of the ion and $N$ is its quantity of bound electrons. $H = d\gamma N^\nu$, where $\nu$ and $\gamma$ are chosen as 0.4 and 1.0 respectively, and $d$ is tabulated for $Z$ values from 2 to 103. $K$ is the kinetic electron energy and $r^2$ is the mean square radius. Finally, $\alpha = 1 - \dfrac{1}{H}$, and

$$F(\alpha) = \sum_{n=0}^{\infty} \frac{\alpha^n}{(n+1)^2}; \; 0 \le \alpha \le 1$$

However, the Bethe formula cannot fit well the stopping power at low velocities due to logarithm in Eq. (14) gives a negative value when its argument is less than one, i.e. $\dfrac{2mv_p^2}{I} < 1$. One way to avoid this is to substitute the logarithm by a set of formulas for high and low velocities interpolated for an intermediate value (Barriga-Carrasco and Maynard, 2005):

$$L_b(v_p) = \begin{cases} L_H(v_p) = \ln\dfrac{2v_p^2}{I} - \dfrac{2K}{v_p^2} & \text{for } v_p > v_{int} \\[2mm] L_B(v_p) = \dfrac{\alpha v_p^3}{1 + G v_p^2} & \text{for } v_p \le v_{int} \end{cases} \tag{18}$$

where

$$v_{int} = \sqrt{3K + 1.5I} \,. \tag{19}$$

Then, stopping power of bound electrons can be calculated by means of (14) and (18):

$$Sp_b(v_p) = \frac{Z_{eff}^2 \, 4\pi n_{at}}{v_p^2} L_b(v_p) \tag{20}$$

$K$ and $I$ can be calculated for the whole plasma ions using Eq. (15) and (16). However, this is a rough approximation, because for a swift charge it is more difficult to ionize the inner electrons than the outer ones. A more detailed calculation can be performed by estimation shell by shell of $K$ and $I$ using a Hartree-Fock method (Barriga-Carrasco and Casas, 2013; Casas et al., 2013), obtaining a better fit to low projectile velocities.

In Eq. (20), stopping power is not a function of temperature and shows a lineal dependence with density. But, it is only valid for plasmas whose ions are not affected by the environment, i.e. that density and temperature can change population and energies of its atomic levels. In these cases, we must to take into account these effects, as lowering potential (Stewart and Pyatt, 1966) where $I$ is reduced by the Coulomb potential of surrounding ions.

In Figure 5 we can see the stopping power calculation using Garbet expressions of $K$ and $I$ for the whole atom, and Hartree-Fock method for theses quantities shell by shell in the same aluminum plasma for proton projectiles.

----------------------------Figure 5--------------------------------

Total stopping power

When an ion beam passes through partially ionized plasma, the stopping power of this medium is estimated by the addition of stopping of bound electrons of plasma atoms to the stopping of free plasma electrons.

In this kind of plasmas the influence of temperature and density is the following: when the temperature rises, the stopping power decreases. On the other hand, when temperature lowers, stopping power increases. This parameter affects mainly to free electrons stopping power. For bound electrons the influence is due to electron level changes, at high temperatures. There is a nonlinear dependence with density, mainly for free electrons. Bound electrons stopping power is a linear function of density, except when there are changes in electron configuration, at high densities.

The dependence with ionization, $q$, is due to the phenomenon called Enhanced Plasma Stopping (Couillaud et al., 1994): When $q$ rises, the stopping power also increases. The reason is the increasing of free electrons density in plasma. It is important mainly at low projectile velocities. For low $q$, bound electrons stopping power is important for medium-high projectile velocities.

In the next section, it will be explained how the energy loss and Bragg peak is calculated from the total stopping power.

**Calculations and results**

Homogeneous WDM

A proton beam loses its energy in a material by means of a continuous and dynamical process: When it initially collides with energy $Ep_i$, is slowed down by means a force, the stopping power, which is a function of proton energy. Its energy decreases, and an equal energy quantity is transferred to the medium. Then the stopping power changes and the proton beam loses a different energy. This process continues until the proton beam traverses all the material or is completely stopped inside.

The easy way to estimate energy losses of a proton beam consists to divide the path inside the plasma into segments connected by nodes where the energy loss is recalculated in each one. In order to do this, we follow the next scheme:

1) Read the input data, where the proton energy is the first column and stopping power the second one.

2) The initial stopping power, $Sp_i$, is obtained by means of an interpolation made between the two nearest stopping values of the input.

3) Estimate the energy loss using this expression:

$$E_{L_i} = \frac{Sp_i}{\Delta x}, \tag{21}$$

where $\Delta x$ is the step length that we use to calculate the stopping inside the plasma.

4) We obtain the new proton beam energy and we continue reevaluating the energy loss in each step, until the proton beam goes outside the plasma or is totally stopped inside.

In Figures 6, and 7 we show, the influence of density and temperature in the range of Bragg peak for homogeneous WDM. A proton beam with an initial energy of 1.5 MeV, passes through a thick partially ionized plasma until is totally stopped.

In the case of density variation, the position of Bragg peak shows a quasilinear dependence on density. In the other case, when density remains constant and temperature is changed, Bragg peak moves its position due to the effects of temperature for low projectile velocities.

----------------------------Figure 6-------------------------------

----------------------------Figure 7-------------------------------

Heterogeneous WDM

Using equations (3) or (11), and (20), it is possible to evaluate the target density profile effects in stopping power calculations for two different density distributions: rectangular shape with a constant density and the piecewise approximation of a trapezium shape with a density profile given by (Andreev et al., 2009):

$$n_i(z) = \frac{2n_{i\,\max}}{1+\exp\left[\dfrac{2x\theta(x)}{l_r} - \dfrac{2x\theta(-x)}{l_{fr}}\right]}. \tag{22}$$

The Eq. (22) is the density distribution obtained when a laser prepulse, with an intensity of $10^{11}$ W/cm$^2$ and a duration of 1 ns, hit a thin target with a thickness of 1 micron or less. Here $x = z - 0.5l_f$ and $\theta(x)$ is a step function. The scale lengths $l_{fr}$ at the front side (in respect to the laser pulse direction), and $l_r$ at the rear side, are obtained from the initial target thickness, $l_f$, $l_{fr}$ and $l_r$ are expressed in microns. For a solid aluminum target, $n_{i\,\max} = 6\cdot10^{22}\,\mathrm{cm}^{-3}$.

We will study a proton beam passing through these plasmas with different densities and lengths, but same superficial densities due to rectangular and piecewise profiles conserve the particle quantity. If the stopping was a linear function of density, it would imply that in both cases proton beam must lose the same energy. But, it was showed in Eq. (12) and Figure 3 that is not linear, although for small differences like here, the stopping power has a quasi-linear dependence on density and for this reason the final proton beam energy is slightly different for both cases. These density profiles and its corresponding energy losses are showed in Figure 8.

----------------------------Figure 8--------------------------------

Application on WDM heating

It is possible to estimate the final temperature of the piecewise plasma profile when a proton bunch passes through it. We suppose a proton beam, made up of $10^{11}$ protons, impacts in a circular area of 50 μm of diameter with an energy of 1.5 MeV. From an initial temperature of 5 eV, the plasma is heated until 25 eV, approximately, with a temperature distribution given by the corresponding piecewise density profiles. This approximation was obtained using an ideal thermodynamic equation of state. We can observe in Figure 9, how the plasma is heated for a proton beam that travels from left to right. In the right part of the plasma, the heating is higher than in the left one, although the density is lower in this area. The reason is that the proton beam has been stopped in the first part, losing its energy and increasing the stopping power as we can see in Figure 7, allowing the bunch to deposit more energy in plasma. Due to plasma density is low in this part, there is more energy to distribute to the ions and a higher temperature is achieved, consequently.

----------------------------Figure 9--------------------------------

**Conclusions**

Two methods have been discussed for the calculation of free electron stopping. The first one uses analytical formulas developed from the linear Vlasov theory and gives fast and accurate results, but only for classical plasmas. The second one is based on the interpolation of a dielectric function valid for plasmas of all degeneracies. Both give very similar results for classical plasmas, but the second one can be also used in stopping power calculations of WDM. Using both methods, the dependence of stopping power with density and temperature in a specific range of classical plasmas has been studied.

Stopping power has been expressed as quasilinear function of density. For this reason the calculations of energy loss give different values for low projectile velocities in the cases of different plasma density profiles. In case of temperature dependencies the stopping power dependence on projectile velocity has been proved important for low velocities and negligible at high velocities, where stopping power graphs tend to be the same value.

For bound electron stopping, two methods have been checked. First, $K$ and $I$ have been deduced for the entire plasma ions by means of expressions obtained by Garbet *et al*. from analytical atomic potentials and solved in the independent particle model. Second, the quantities $K$ and $I$ are estimated in every atomic shell using a Hartree-Fock method. Both quantities are used in an interpolation that divides the stopping power graph in two parts. These are the low velocity stopping power with a realistic behavior in the proximities of zero, and at medium-high velocities asymptotically the Bethe formula result is approximated. The Hartree-Fock method results in more accurate energies for low $Z$ elements, although Garbet *et al*. expressions are useful for medium-high $Z$ elements.

WDM has been analyzed for homogenous and heterogeneous cases. For the first one, the stopping range and the position of Bragg peak has been estimated by means of an iterative method that estimates dynamically the energy loss of the projectile inside the plasma. It has been found that the dependency on density and temperature is the same as in the previous section: This holds for density. Also the Bragg peak position has shown a quasilinear dependence on density. In regard to the temperature the Bragg peak position has changed for low proton energies.

For the heterogeneous case, a WDM density profile, given by an analytical expression, has been studied. A continuous profile of density has been approximated using a rectangular profile and a piecewise function, conserving the particle quantity. A slightly difference in the energy losses by the proton has been found due to the quasilinear effect of density in stopping power. Finally, using the same piecewise profile, the plasma heating until WDM conditions has been estimated as function of density profile and stopping power. An overall heating of 20 eV has been obtained, but with small differences due to plasma density profile, projectile velocity and stopping power.

## Acknowledgements

We are grateful to Junta de Comunidades de Castilla-La Mancha for the Programa Operativo Regional FSE 2007/2013. The research leading to these results has received funding from Deutsche Forschungsgemeinschaft within the program CRC/Transregio

18 and from LASERLAB-EUROPE (grants agreement n° 284464, EC's Seventh Framework Program)

**Figure Captions**

**Fig. 1.** (Color online) Temperature-density diagram divided in two regions by the blue line, and dashed dark green. Proton projectiles are considered. Upper zone is classical plasma region while lower one is the degenerate region. The red box delimited the Warm Dense Matter (WDM) conditions.

**Fig. 2.** (Color online) Stopping power as a function of the projectile energy (protons). Blue dashed line calculated by Eq. (3). Read dotted line estimated by interpolation of RPA stopping database. Green dash-dotted line direct calculation of RPA DF by Eq. (12).

**Fig. 3.** (Color online) Ratio of stoppings for different densities as a function of proton energy. The reference stopping is the corresponding to $1 \cdot 10^{19}$ e-/cm3. Dashed lines: Estimated using RPA. Dotted lines: Calculated by means of analytical formula.

**Fig. 4.** (Color online) Ratio of stoppings for different temperatures as a function of proton energy. The reference stopping is the corresponding to 50 eV. Dashed lines: Estimated using RPA. Dotted lines: Calculated by means of analytical formula.

**Fig. 5.** (Color online) Stopping power of bound electrons as function of proton energy. Two different sets of K and I for the aluminum ion plasma are used in the calculation of stopping power using Eq. (15) (Bethe) and (21) (Barriga-Carrasco & Maynard).

**Fig. 6.** (Color online) Bragg peak position as function of density. Depth where proton beam is totally stopped: Black dash-dotted line: 0.182 cm. Blue dashed line: 0.0217 cm. Red dotted line: 0.00281 cm.

**Fig. 7.** (Color online) Bragg peak position as function of temperature. Depth where proton beam is totally stopped: Black dash-dotted line: 0.182 cm. Blue dashed line: 0.166 cm. Red dotted line: 0.1655 cm.

**Fig. 8.** (Color online) The target density profiles (top) and its corresponding energy loss functions (bottom). The final energy was 1.409 MeV for the rectangular profile and 1.401 MeV for the piecewise profile.

**Fig. 9.** (Color online) Red short dot line represents the plasma temperature after the heating by proton beam. Dark solid line is the piecewise profile density that approximates Eq. (23). Initial plasma conditions are the same as Fig. 8.

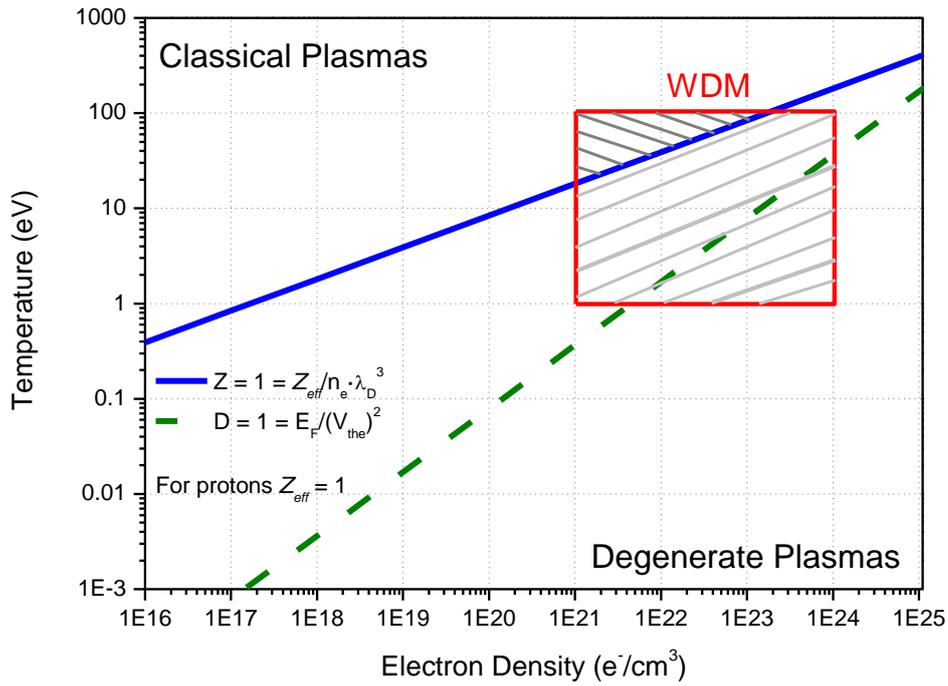

Fig. 1. David Casas, Alexander A. Andreev, Matthias Schnürer, Manuel D. Barriga-
Carrasco, Roberto Morales and Luis González-Gallego

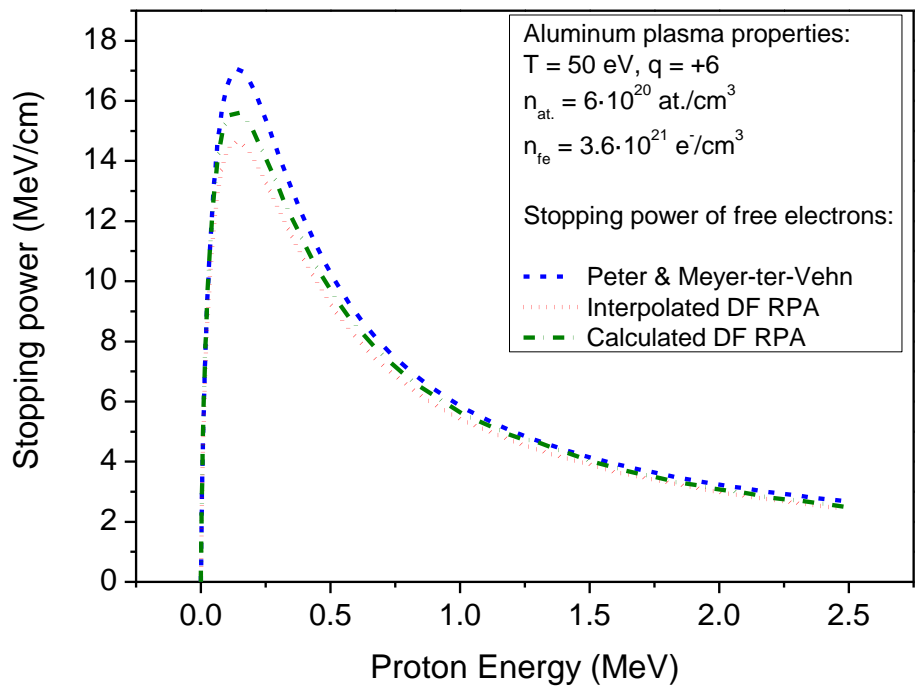

Fig. 2. David Casas, Alexander A. Andreev, Matthias Schnürer, Manuel D. Barriga-
Carrasco, Roberto Morales and Luis González-Gallego

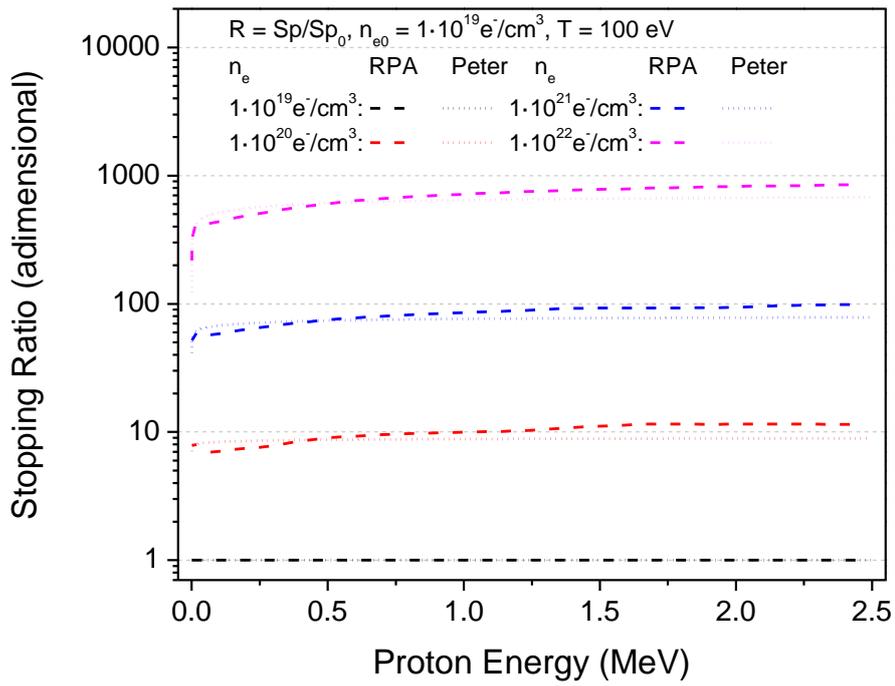

Fig. 3. David Casas, Alexander A. Andreev, Matthias Schnürer, Manuel D. Barriga-Carrasco, Roberto Morales and Luis González-Gallego

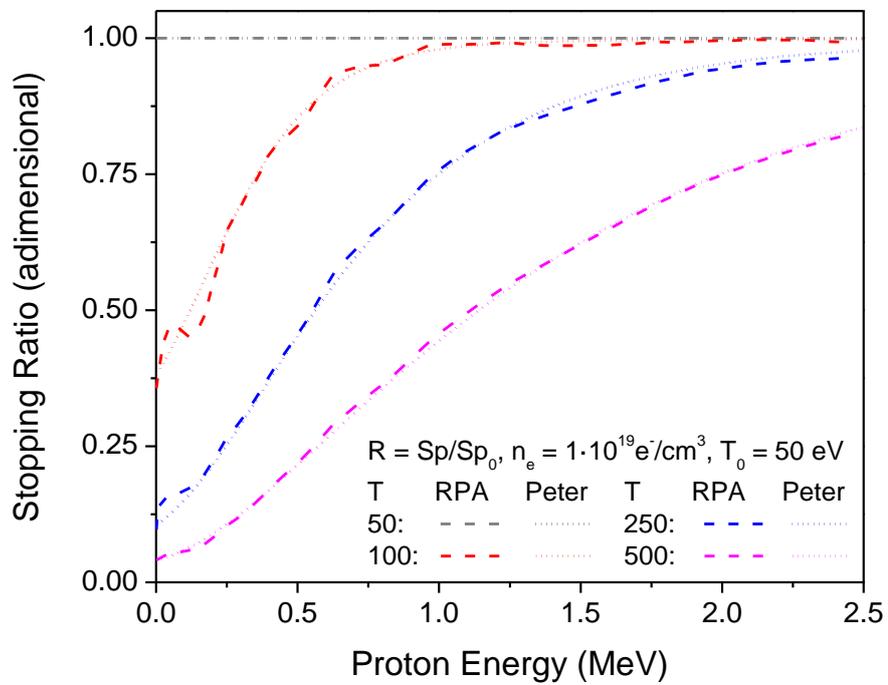

Fig. 4. David Casas, Alexander A. Andreev, Matthias Schnürer, Manuel D. Barriga-Carrasco, Roberto Morales and Luis González-Gallego

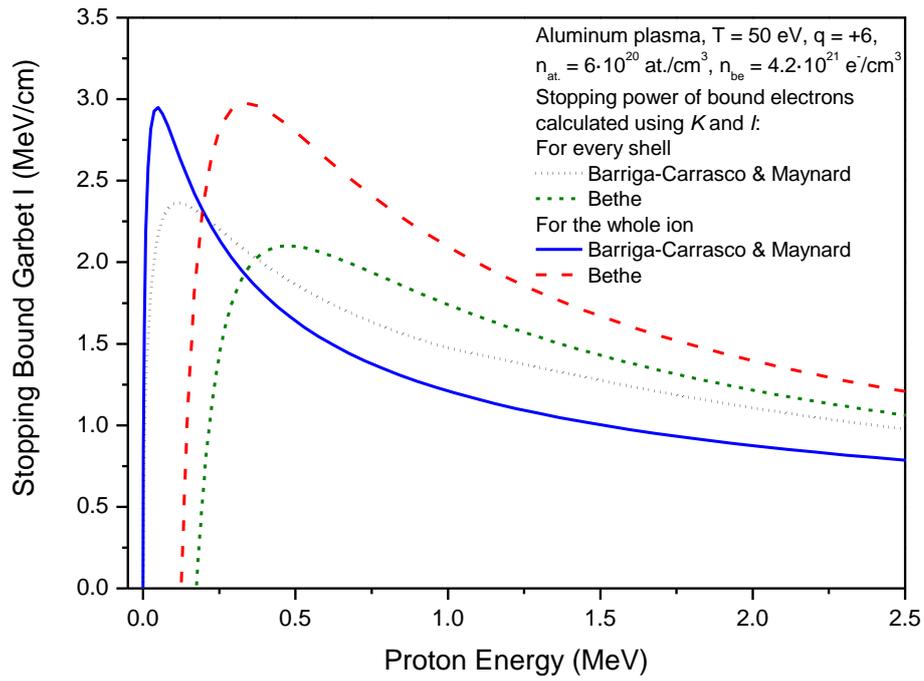

Fig. 5. David Casas, Alexander A. Andreev, Matthias Schnürer, Manuel D. Barriga-Carrasco, Roberto Morales and Luis González-Gallego

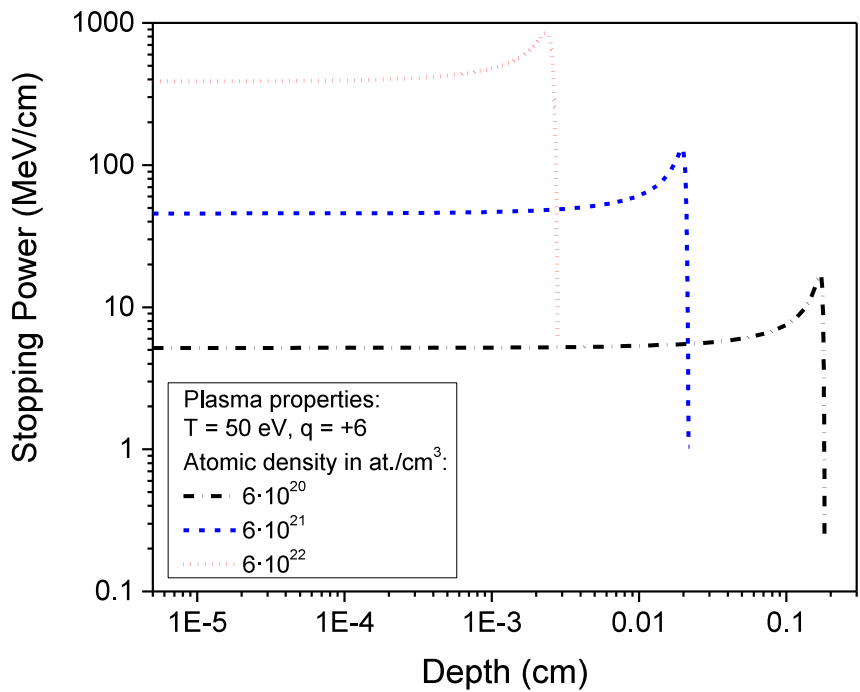

Fig. 6. David Casas, Alexander A. Andreev, Matthias Schnürer, Manuel D. Barriga-Carrasco, Roberto Morales and Luis González-Gallego

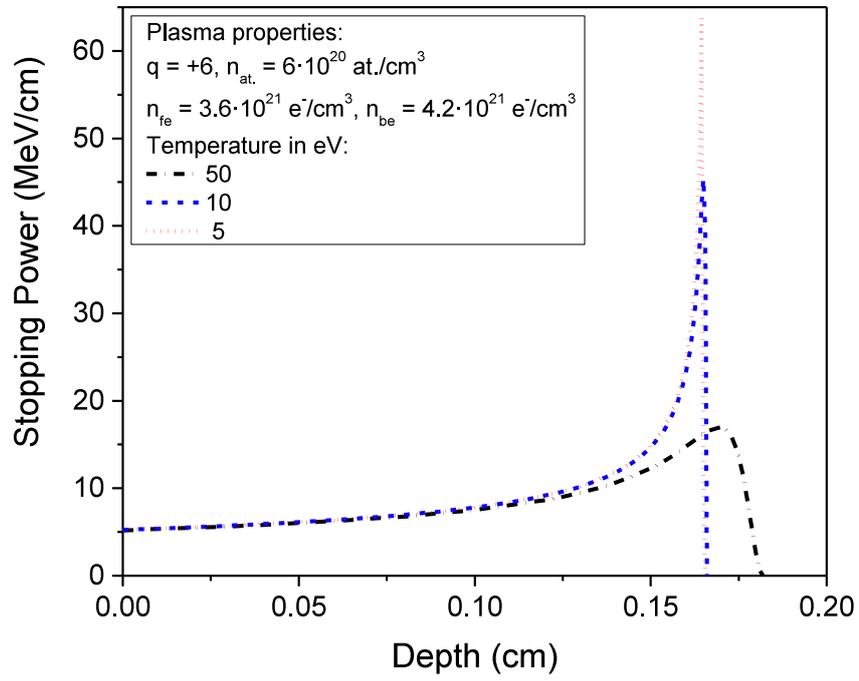

Fig. 7. David Casas, Alexander A. Andreev, Matthias Schnürer, Manuel D. Barriga-
Carrasco, Roberto Morales and Luis González-Gallego

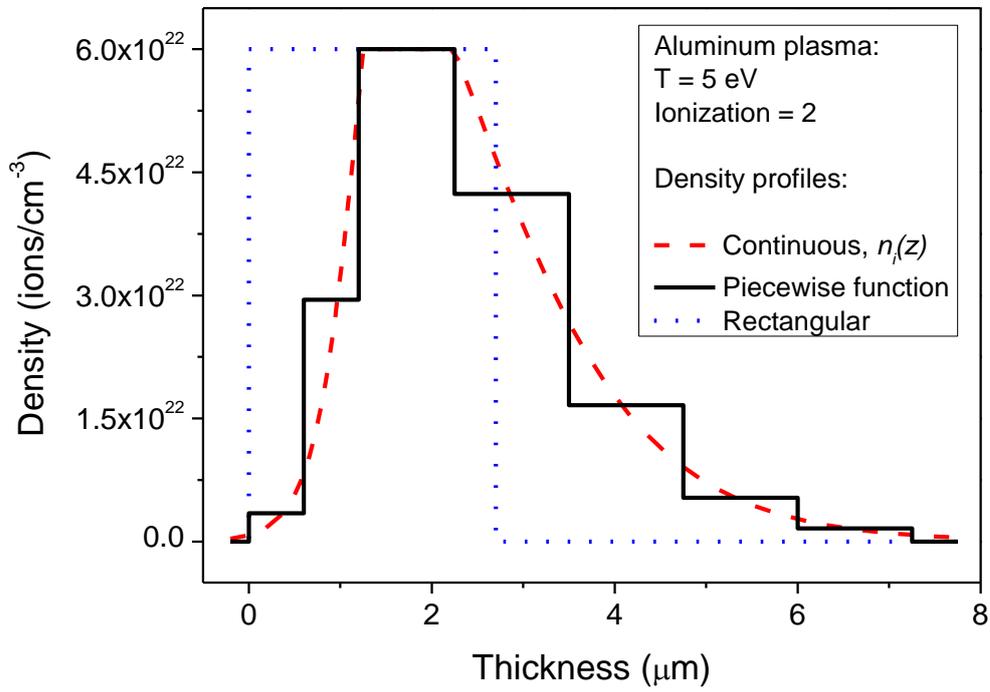

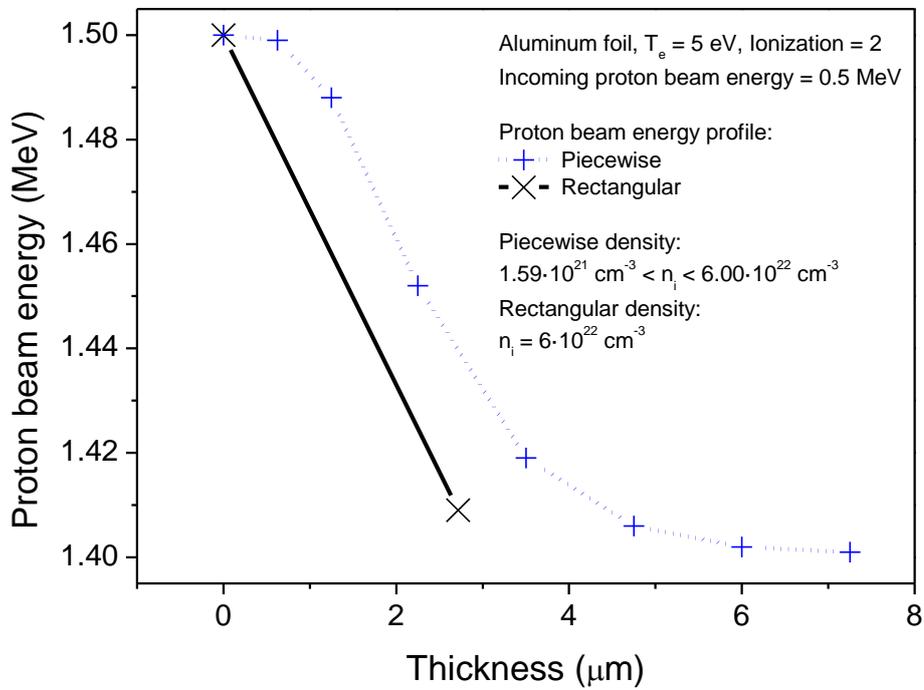

Fig. 8. David Casas, Alexander A. Andreev, Matthias Schnürer, Manuel D. Barriga-Carrasco, Roberto Morales and Luis González-Gallego

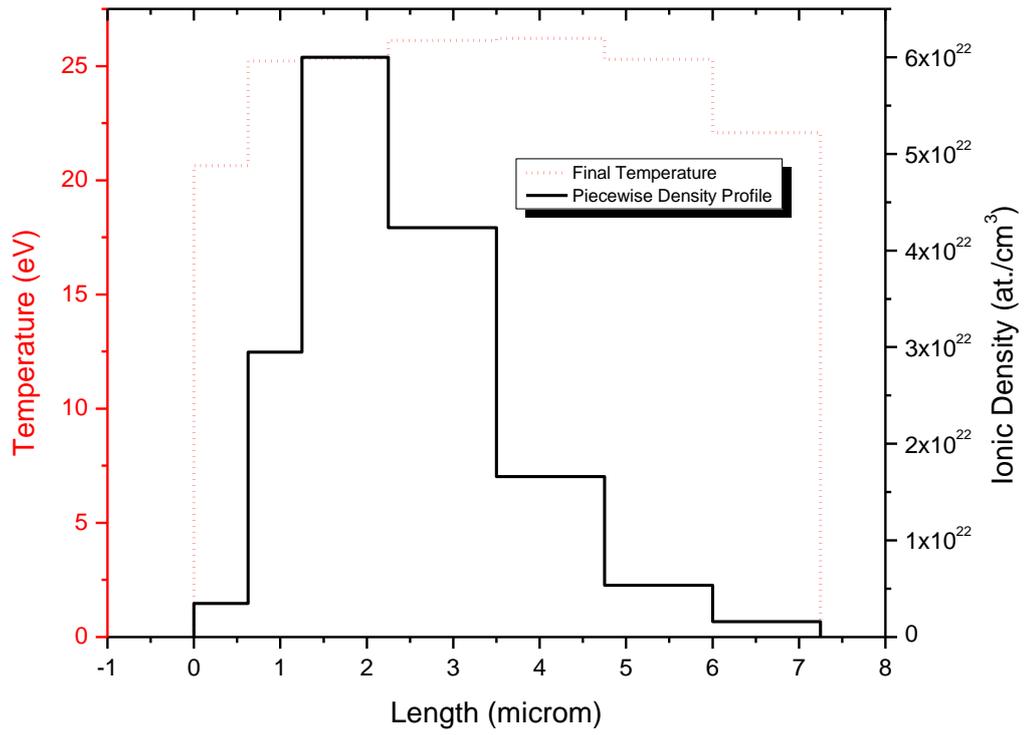

Fig. 9. David Casas, Alexander A. Andreev, Matthias Schnürer, Manuel D. Barriga-Carrasco, Roberto Morales and Luis González-Gallego